\documentclass[11pt,a4paper]{article}

\usepackage[utf8]{inputenc}
\usepackage[T1]{fontenc}
\usepackage{lmodern}
\usepackage{geometry}
\usepackage{setspace}
\usepackage{amsmath,amssymb,amsthm,mathtools}
\usepackage{braket}
\usepackage{subcaption}
\usepackage{bm}
\usepackage{graphicx}
\usepackage{hyperref}
\usepackage{cite}
\usepackage{enumitem}
\usepackage{tensor}
\usepackage{slashed}
\usepackage{bbm}
\usepackage{tikz}
\usepackage{authblk}
\usetikzlibrary{arrows.meta, decorations.pathreplacing}

\hypersetup{
	colorlinks=true,
	linkcolor=blue,
	citecolor=blue,
	urlcolor=blue
}
\newcommand{\xb}{\overline{x}}

\title{Conjugate measurements, equilibration and emergent classicality}    
\author{Adarsh S}
\author{P N Bala Subramanian}
\author{T P Sreeraj}
\affil{Department of Physics,  National Institute of Technology Calicut,
	Kozhikode - 673601, Kerala, India\\
	\texttt{adarsh\_p210009ph@nitc.ac.in, pnbala@nitc.ac.in, sreerajtp@nitc.ac.in}}
\begin{document}

		\maketitle
	\begin{abstract}
		Simultaneous decoherence of conjugate observables of an open quantum system leads to 
		%		equilibration of the system. This leads to 
		a classical statistical mechanical description  with constant phase space probability density in terms of a uniform ensemble. 
		%		with equal apriori probability. 
		We investigate a scenario where this may be realized by measurement of basic conjugate observables of a quantum system by the environment. \nocite{*}%% Remove this line from your manuscript.
	\end{abstract}
	
	
%	\begin{highlights}
%		\item 
%		\item 
%		\item 
%	\end{highlights}
%	
%	
%	%\nocite{*}
%	
%	% Keywords
%	% Each keyword is seperated by \sep
%	\begin{keywords}
%		
%		Equilibration \sep Quantum decoherence  \sep  Quantum measurement
%	\end{keywords}

	\section{Introduction}
	%	Systems equilibrate and
	%%	. Quantum systems
	%	 decohere. 
	%	When a large class of systems show two kinds of generic behavior, it is usually profitable to investigate whether there is any connection between the two. 

	The postulate of equal apriori probability for isolated systems serves as a starting point from which the edifice of equilibrium statistical mechanics is traditionally established. System under interest is thought of as a subsystem of an isolated system and an equilibrium probability distribution of its microstates derived under various restrictions on the subsystem.
	%leading to various kinds of ensembles. 
	In the classical regime, there have been various approaches to justify equal apriori probability  hypothesis based on ergodicity where attempts were made to show that a closed system under time evolution spends equal amount of time in equal volume of phase space. Chaos is generally considered as an important contributing ingredient. As early as 1929, von Neumann \cite{VNergodicity} recognized that the problem is more tractable for quantum systems. For quantum systems, one of the approaches used to justify equal apriori assumption is quantum typicality, were it is shown that \cite{popescu,goldstein} for almost all states of a large system, the reduced density matrix of a small subsystem is essentially equal to the reduced density matrix that one gets if one had taken the large system to be in a maximally mixed state, i.e, if one had assumed equal apriori postulate for the large system. However, for a complete proof of the postulate one has to show that quantum dynamics inevitably leads to such a state. A related approach is where one shows \cite{ETH} that for observables of an isolated system obeying Eigenstate Thermalisation Hypothesis, the infinite time average of an observable turns out to be equal to that computed in a micro canonical ensemble. Many non-integrable systems have been shown to satisfy ETH. Notable exceptions to ETH are integrable systems and systems with quantum scars or showing many body localisation. One can ask what features of the (quantum) dynamics of the system leads to ETH and equal apriori probabilties. 
	
	%Apart from the above top-down approaches, 
	An alternative 
	%bottom up 
	approach was that of Jaynes \cite{jaynes} where subjectivity of lack of information about the state of the system plays the main role. In such an approach a maximal lack of information forces one to assume that all states are equally probable, leading to what is called a uniform ensemble. One further puts constraints on the system based on new information gained and finds the equilibrium probability distribution of microstates by maximizing the lack of information (entropy) consistent with the constraints. 
	For instance, a constraint of fixed energy and particle number leads to a microcanonical ensemble and a  
%	For instance, 
	a constraint of fixed particle number and ensemble average of energy leads to a canonical distribution.
	%A constraint of fixed ensemble average of energy and particle number leads to grand canonical ensemble etc.
	In quantum systems, loss of information fundamentally occurs in the form of entanglement of the system state with inaccessible parts of the environment, leading in general to a mixed reduced density matrix. One can ask what features of the (quantum) dynamics of an open system leads to a uniform ensemble with equal apriori probability.
	%The question of what features of the interaction of the system with the environment leads to a maximally mixed state and hence equal apriori probability for an open unconstrained subsystem is an important one. 
	Since, entanglement of the subsystem with the environment is also the root cause of quantum decoherence of the subsystem, the issue of equilibration becomes related to the issue of quantum decoherence. In this paper, we will show that a simultaneous measurement of conjugate observables of the system by the environment, leads to decoherence of conjugate observables which inevitably lead to equal apriori probability for all the states of the system. 
	
	%there is an alternate bottom-up approach \cite{jaynes}, where one considers a completely open system without any constraints and assume equal apriori probability for any state. One can further impose constraints on top of it and derive probability distributions by maximizing entropy 
	
	%If one further assumes that the subsystem settles down to a time independent, equilibrium distribution,   

	%************************************

	%There are two generic behaviours that emerge from quantum dynamics of systems that are of physical interest:
	% one is 
	%decoherence and 
	% the other is 
	%equilibration. Even though on first look, these behaviours looks very different, both can be attributed to the behavior of reduced density matrix of the system when it interacts with the environment. 
	%There are systems where both these features are simultaneously present, either one is present. There are also systems which can be attributed neither of these properties (may be like superconducitity or something), but then they have to be treated in their full quantum mechanical technicality, which in-principle we know how to describe. It is in systems where there is an emergence of these attributes, while being fundamentally described using quantum mechanics, where the questions are interesting. A macroscopic description of the world that we know of, called Classical, appears in this context as an emergent behaviour. 
	\section{Equilibration, open quantum systems and quantum decoherence}
	Equilibrium is typically defined as follows:  If system 
	%	wait for a time $T_0$  where all 
	observables averaged over a spatial scale $\chi$ and a time scale $\tau$
	%	 (resolving power of the measurement apparatus) 
	does not change appreciably 
	(i.e, not more than the precision $\delta o$ of  measurement of an arbitrary observable $O$) 
	over a time scale $T$
	%	  (time scale in which the entire experiment is done)
	, and is independent of the initial state, then the system is said to be in equilibrium. $\chi$ and $\tau$ give the spatial and temporal resolution of measurement apparatus and $T$ is the time scale in which the entire experiment is done. Once the system is disturbed out of equilibrium, one typically has to wait for a time $T_0$ for the system to ease back to equilibrium. The definition depends on the scales $\chi, \tau$ and $T$ as well as the measurement precision $\delta o$  which are not features inherent to the system but depends on the observer. Therefore, the same situation may count as equilibrium or not depending on the observer.  Such subjectivity of definition is also true for classicality. The same system may look classical or quantum mechanical depending on the scales involved in measurement.
	
	Consider an open quantum mechanical system $S$ interacting with an environment ${\cal E}$. The state of (system + environment), $S\mathcal{E}$  at time $t$ is $|\Psi(t)\rangle = \sum\limits_{i,j} c_{ij}(t) |s_i\rangle |e_j\rangle=\sum\limits_{i} |s_i\rangle |\tilde{e}_i(t)\rangle,$ where $|\tilde{e}_i(t)\rangle=\sum\limits_{j}c_{ij}(t) |e_j\rangle$, and $|s_i\rangle$ and $|e_j\rangle$ are orthonormal bases of system and environment Hilbert spaces. Time average of a coarse grained system observable $O$ over a time $\tau$ at time $t'$ is , 
	%	\begin{widetext}{2}
	%\begin{strip}
	\begin{align}
	\bar{O}(t')&=\frac{1}{\tau}\int\limits_{t'}^{t'+\tau} \langle \Psi(t)|O|\Psi(t)\rangle~ dt=\sum_{i,j} \langle s_i|O|s_j\rangle \Big[\frac{1}{\tau}\int\limits_{t'}^{t'+\tau}{d}_{ij}(t) ~dt\Big]=\frac{1}{\tau}\int\limits_{t'}^{t'+\tau}Tr(\rho_s(t) O) ~dt\nonumber\\&= Tr(\bar{\rho}_s(t')O)
	\label{avgo}
	\end{align}
	%\end{strip}
	%\end{widetext}
	where, $t'\geq T_0$,  
	%	$O$ is a system observable averaged over $\chi$ and
	${d}_{ij}(t)= \langle \tilde{e}_i(t)|\tilde{e}_j(t)\rangle  = \langle s_j|\rho_s| s_i\rangle $ and $\rho_s=Tr_{\mathcal{E}}(|\psi\rangle \langle \psi|)= \sum_{i}\langle e_i|\psi(t)\rangle \langle \psi(t)|e_i\rangle$ is the reduced density matrix. Given a fixed   $\chi, \tau$ and $T$, for the system to be in equilibrium the following conditions have to be satisfied: 
	\begin{enumerate}
		\item The time average of $d_{ij}$ i.e, the matrix elements of the reduced density matrix  in any (and therefore all) basis, over time $\tau$ becomes effectively time independent \footnote{If the observable $O$ changes with time explicitly due to interaction with the environment, Time average of $Tr(\rho_s O)$ can become time independent without $\rho_s(t)$ becoming time independent. We don't consider such observables here.} such that  $\bar{O}$ does not change by more than the measurement precision $\delta o$. This means, one can always find an equilibration time $T=T(\delta o)$ for any observable.  
		%	 	 I.e, \begin{align}
		%	 		\lim\limits_{t'\rightarrow \infty}\frac{d}{dt'} \bar{\rho}_s(t') =0
		%	 	\end{align} 
		\item $\bar{\rho}_s(t')$ lose all memory of the initial state of the system. 
	\end{enumerate}
	In general,  $\bar{\rho}_s$ would be a mixed state.
	%and Von Neumann entropy of this state is the entanglement entropy. 
	%	  In the second equality, we have used the fact that $\langle s_j|\rho_s| s_i\rangle=\langle \tilde{e}_i(t)|\tilde{e}_j(t)\rangle$.
	%	 is the matrix elements of reduced density matrix $\rho_s(t)$ in the system basis.   		
%	The standard formulation of equilibrium quantum statistical mechanics hinges on two assumptions: (a) The assumption of random phases and (b) Assumption of equal a priori probability. 
%	%	   For an open system S without any constraints, in equilibrium, 
%	Assumption (a) means that off diagonal elements of $\bar{\rho}_s(t')$(in any basis) are zero and    (b) means that the diagonal elements become time independent and constant. Assumptions (a) and (b)  implies ( and are stronger than) the conditions for equilibrium stated above.
	
	An isolated  quantum mechanical system described by a wavefunction would preserve its  \emph{quantum}-ness on time evolution. However, in a realistic view, the quantum mechanical system is continuously interacting with the environment which leads to the entanglement of system state with that of the environment\cite{zeh}\cite{Zurek1}\cite{schlos}. 
	%Quantum decoherence \cite{zeh}\cite{Zurek1}\cite{schlos} posits that 
	As far as moments of system observables  and therefore their probability distributions are concerned, one can replace the total density operator with the reduced density operator.
	If the reduced density matrix written in an eigenbasis of a system observable $O$ is diagonal, the observable is said to have decohered as this reduced density matrix doesn't give quantum interference effects of the observable. 
	%Since, , such quantum interference become experimentally unobservable.
	One expect that, in a quantum to classical transition, all observables, and conjugate observables in particular, decoheres. That is, the reduced density matrix becomes diagonal in eigenbasis of conjugate observables. However, this is only possible if the reduced density matrix is a constant multiple of identity. This in turn implies that all eigenvalues of any non-degenerate observable will be equally probable and the probability of degenerate observables becomes proportional to the degeneracy factors. This indicate that the emergence of classicality in the above sense implies equal apriori probability. We now study such a scenario for an open quantum system were environment can be thought of as (imprecisely) measuring conjugate observables of the system. We show that for the case where interactions are dominant, the reduced density matrix becomes diagonal in the eigenbasis of all observables and therefore represents a uniform ensemble.
	The probability distributions of conjugate observables become independent of each other leading to 
	%	 thereby leading to realization of both these notions of classicality. 
	%	 This further leads to a 
	classical statistical description with
	%	 based on a classical density matrix with 
	constant probability density in phase space.  

	\section{Conjugate measurements, classicality and equal apriori probabilities.}
	\label{imperfect}
	% Consider a quantum system ${\cal S}$ which is interacting with an environment ${\cal E}$. Let's call the composite system ${\cal S}+{\cal E}$  as the universe $U$. 
	The Hamiltonian of an open quantum system can be written as
	%  e is given by :
	\begin{align}
	H=H_0+H_E +H_{int} 
	\end{align} 
	where $H_0$ is the system Hamiltonian, $H_E$ is the environment Hamiltonian and $H_{int}$ describes the interaction between the system and the environment. Let us further assume that the environment has two independent, non-interacting, degrees of freedom   
	% The environment Hilbert space is further assumed to be a tensor product of two subspaces corresponding to two different degrees of freedom of the environment 
	which we denote as $\hat{p_1}$ and $\hat{y}_2$ which couples to the generalized position $x$ and the corresponding conjugate momentum $p$ of the system. We call these as environment 1 and environment 2. The conjugate variables corresponding to $\hat{p_1}$ and $\hat{y}_2$ are denoted as $\hat{y}_1$ and $\hat{p}_2$. Following Von Neumann \cite{VNergodicity,VNold}, one can assert that the simultaneous imprecise measurement of the physical observables $x$ and $p$ of the system, is actually performed by measuring the conjugates $\hat{y}_1$ and $\hat{p}_2$.
%	 Measurement of $\hat{y}_1$ and $\hat{p}_2$ can be done with arbitrary precision as they commute. 
However, these values only reflect $x$ and $p$ imprecisely. 
%in accordance with uncertainty principle (see eqn ***).    
	% of the environment variables $\hat{p_1}$ and $\hat{y}_2$ which couples with the system. 
	We take both these couplings to be of the Von Neumann form \cite{VNold,TS,kelly} leading to an interaction Hamiltonian $H_{int}$ of the form: 
	\begin{align} 
	\label{hint}
	\hspace{-1cm}H_{int} &= \int dx |x\rangle \langle x| \otimes x \hat{p}_1 \otimes \hat{\mathbf 1} + \int dp |p\rangle\langle p| \otimes \hat{\mathbf 1} \otimes p \hat{y}_2 = \hat{x}\hat{p}_1+\hat{p} \hat{y}_2
	\end{align}
	
	In the above expression, anything without a hat are numbers. The first term in the $H_{int}$ above is the form of interaction Hamiltonian for standard von Neumann \cite{VNold} quantum measurement model for the measurement of $x$. The second term is added so that $x$ and $p$ are treated in the same footing.
	%    corresponding to the measurement of $\hat{x}$ and $\hat{p}$ by Environment 1 and 2. 
	%One of the distinguishing features of a quantum system when compared to a classical system is quantum interference. In practice it is hard to realize and maintain quantum interference effects.
	% Vanishing of quantum interference effects in the measurement of an observable in open quantum systems can be understood through the mechanism of quantum decoherence. The environment acts like an observer and the interaction of the system with the environment acts as a entangling process by which the system state entangles with the environment.
	Assuming that $H_{int}$ dominates in the Hamiltonian, if we start with a separable state of the system + environment, the corresponding density operator  evolves with time as follows: 
	%\begin{multicols}{2}
	\begin{align} 
	\hspace{-.6cm}\rho=|\psi\rangle |E_1\rangle |E_2\rangle \langle \psi| \langle E_1|\langle E_2| \xrightarrow[evolution]{time}\rho(t)= |\Psi_f\rangle \langle \Psi_f|
	\end{align}
	where,
	\begin{align}
	|\Psi_f(t)\rangle&= e^{-\frac{i}{\hbar}t H_{int}} |\psi\rangle|E_1\rangle|E_2\rangle= \int dx dy_1dy_2 ~\psi(x) E_1(y_1) E_2(y_2) | x+y_2t\rangle~ |y_1+xt+\frac{1}{2} y_2t^2\rangle~ |y_2\rangle ;\nonumber\\\psi(x)&=\langle x|\psi\rangle ~;~ E_1(y_1)=\langle y_1|E_1\rangle~;~ E_1(y_1)=\langle y_2|E_2\rangle
	\end{align}
	%\end{multicols}{2}
	The full statistical information about local observables is contained in the reduced density matrix, as replacing the full density matrix with a reduced density matrix gives the same value for all moments of any local observable. Therefore, in order to see how the system looks to a local observer who only has access to local observables, we trace out the environment and find out the reduced density matrix. 
	%\begin{multicolst}{2}
	\begin{align}
	\hspace{-2cm}\rho_s(t)&= Tr_{\cal E}\rho(t) \nonumber\\&= \int dx d\bar{x} ~dy_1 dy_2~ \psi(x) \psi^*(\bar{x})~ E_1(y_1)E_1^*\big(y_1+(x-\bar{x})t\big)~ |E_2(y_2)|^2 ~|x+y_2t\rangle \langle \bar{x}+y_2t|
	\end{align}
	%\end{multicols}{2}
	Since, $E_1(y_2)$ is a "nice" square integrable function \footnote{We ignore pathological functions where $\int |\psi(x)|^2 dx < \infty  \implies \lim_{|x|\rightarrow \infty} \psi(x)  \neq 0$},  given enough time, 
	$\int dy_1 E_1(y_1) E_1^*(y_1-(\bar{x}-x)t)$ vanishes unless $\bar{x}=x$. Therefore, after a sufficient amount of time, the off diagonal elements of $\rho_s(t)$ are suppressed and it tends towards a diagonal matrix. This signals an inability of the reduced density matrix to represent linear combination of $x$ states. After a long time the reduced density matrix becomes: 
	%\begin{multicols}{2}
	\begin{align}
	\rho_s(t)&=\int dx  dy_2 |\psi(x)|^2 |E_2(y_2)|^2 | x+y_2t\rangle \langle x+y_2t| \nonumber\\&= \int dX dy_2 |\psi(X-y_2t)|^2 |E_2(y_2)|^2 |X\rangle \langle X|
	\label{xreddensity}
	\end{align} 
	%\end{multicols}{2}
	where, $X= x+y_2t$.
	Now, if we compute $\rho_s(t)$ in the conjugate momentum basis, we get 
	\begin{align}
	\rho_s(t)&= \int dp d\bar{p}~ dp_1 dp_2 ~\psi(p)\psi^*(\bar{p})~|E_1(p_1)|^2 E_2(p_2)E_2^*\big(p_2-(p-\bar{p})t\big) ~|p-p_1t\rangle\langle \bar{p}-p_1t|
	\end{align}
	As before, because $E_2(p_2)$ is square integrable, $\rho_s(t)$ becomes diagonal after sufficient amount of time and maybe written as :
%	 Therefore, after the system couples with the environment for a sufficiently long time, the reduced density operator may be written as 
	\begin{align}
	\rho_s(t)&= \int dp dp_1 |\psi(p)|^2 |E_1(p_1)|^2 |p-p_1t\rangle \langle p-p_1t | \nonumber\\&= \int dP dp_1 |\psi(P+p_1t)|^2 |E_1(p_1)|^2 |P\rangle \langle P|
	\label{preddensity}
	\end{align}
	
	From (\ref{xreddensity}), it is clear that $\rho_s$ represents a statistical mixture of position eigenstates 
	% and not of a linear combination of $|x\rangle$ states. However,
	and from (\ref{preddensity}), it is a statistical mixture of momentum eigenstates. The probability distributions of $x$ and $p$ seems to have become independent from each other signaling classical behavior.
	% This implies that momentum and position eigenstates can't have non zero overlap.  
	Moreover, $\lim\limits_{x\rightarrow \infty}\frac{d}{dx}|\psi(x)|^2 =0$ implies that the reduced density operator in $x$ as well as $p$ basis becomes effectively time independent at large times. 
	$P(x)= \rho_s(x,x)$ and $\bar{P}(p)= \rho_s(p,p)$ gives probability distribution of position and momentum respectively.   $\rho_s$ in the limit of large time is 
	\begin{align}
	\rho_s &= \int dx P(x) |x\rangle \langle x| \nonumber\\&= \int dx dp dp' P(x) e^{-\frac{i}{\hbar}(p-p')x} |p\rangle \langle p'|\nonumber\\&=  \int dp \bar{P}(p) |p\rangle \langle p|  
	\end{align}
	This implies
	\begin{align}
	\int P(x) e^{-\frac{i}{\hbar}(p-p')x} dx = \delta(p-p') \bar{P}(p)
	%	 \implies P(x)=\bar{P}(p)\xrightarrow{t\rightarrow \infty} \mathrm{constant}=0 IS THIS THE ONLY WAY THIS CAN HAPPEN?????
	\end{align}
	This is consistent with the fact that when time is sufficiently large $P(x)$ as well as $P(p)$ evens out and becomes independent of $x$ and $p$ and becomes a constant $C$, which tends towards zero as $Tr\rho_s=1$. The joint distribution function $P(x,p)$ will be $P_x(x)P_p(p)$. This represents a statistical system with constant phase space density. At this point, it should be mentioned that even though in our analysis we have assumed $x$ and $p$ to have infinite range for simplicity, a more rigorous analysis should keep $x$ and $p$ bounded (say by considering the system to be a lattice) so that a uniform distribution doesn't vanish. This is an implicit assumption whenever we talk about equal apriori probability as a uniform distribution over an infinite range is, strictly speaking, not defined. Off course, boundedness of $x$ and $p$ is a reasonable assumption for any real system. 
	
	For a system consisting of non interacting free particles, the above behavior of the reduced density matrices doesn't change even if we don't ignore the action of the system Hamiltonian. For a system consisting of a free particle, $\rho_s$ is given by : 	
%		\begin{strip}
		\begin{align}
		\rho_s(t)&= \int dp d\bar{p}\Big(\int dp_1 dp_2 |E_1(p_1)|^2 E_2(p_2) E^*(p_2+(p-\bar{p})t) \psi(p+p_1t)\psi^*(p+p_1t) \nonumber\\ &e^{\frac{-i}{\hbar} \big[\frac{(p^2-\bar{p}^2)}{2m}t + \frac{(p-\bar{p})}{2m}p_1t^2\big]} \Big) |p\rangle \langle \bar{p}|
		\label{redh0}
		\end{align}
%	\end{strip}
			It is clear from above that after sufficiently long time, $\rho_s$ is diagonal, time independent and constant in p basis. This in turn implies the same applies in x basis also.  
			
%			The behavior of the reduced density matrix seems to be different for various interaction strengths. If the interactions are so weak that it can be ignored, the reduced density matrix is not diagonal and the distribution along the diagonal is not uniform and depends crucially on the initial state. However, averaging over  This is the regime where Ehrenfest theorem implies that classical trajectories emerge\cite{messiah} If the interactions dominate, 
		%
		%***************************************
		%
		%The reduced density matrix for the system in the position basis is
		%\begin{align}
		%	\rho_s(t) = \int dx \,d\xb\, dy_1\, dy_2\, \psi(x)\psi^*(\xb) E_1(y_1)E^*_1(y_1+(x-\xb)t) |E_2(y_2)|^2 \ket{x+y_2 t}\bra{\xb+y_2 t}
		%\end{align}
		%and in the momentum basis
		%\begin{align}
		%	\rho_s(t) = \int dp\, d\pb\, dp_1\, dp_2 \psi(p)\psi^*(\pb) |E_1(p_1)|^2 E_2(p_2)E_2^*(p_2-(p-\pb)t) \ket{p- p_1 t}\bra{\pb -p_1 t}
		%\end{align}
		%in which the tildes have been omitted that represent that we are looking at the appropriately Fourier transformed versions of the wavefunctions in the previous equation.
		\section{An illustrative example}
		\begin{figure*}[t]
			\centering
			\begin{subfigure}{\textwidth}
				\includegraphics[width=1\textwidth]{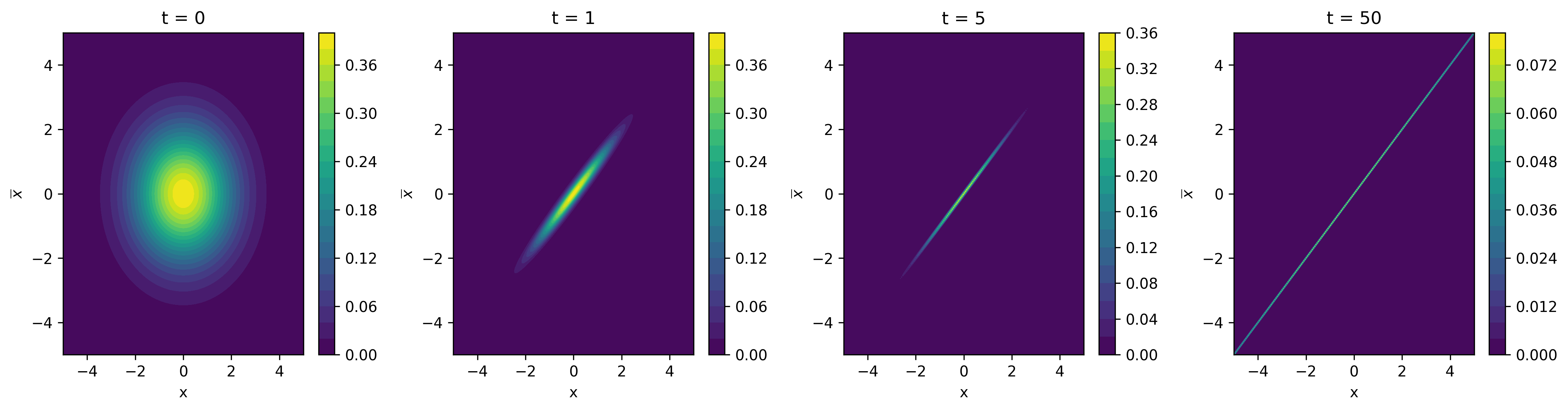}
			\end{subfigure}
			\begin{subfigure}{\textwidth}
				\includegraphics[width=1\textwidth]{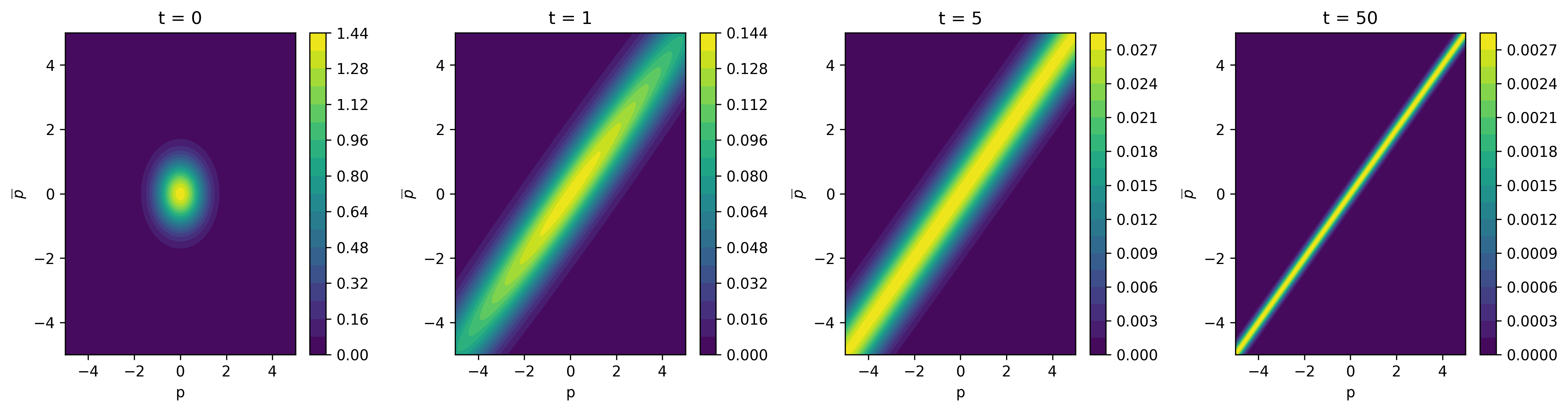}
			\end{subfigure}
			\caption{Graphical depiction of the time evolution of reduced density matrix for a particle with environment interaction of the form (\ref{hint}).  Initial wavefunctions of the system and both the environments  are chosen to be Gaussians with standard deviation of $1$. Units are chosen such that $\hbar$ is set to 1. The contour plot (a) and (b) shows the time evolution of $|\rho_s(x,\bar{x})|$ and  $|\rho_s(p,\bar{p})|$  respectively.}
		\end{figure*}
		
		As a realistic example of the setup, let us look at a case where the system and environment wavefunctions are Gaussians. For convenience, we will set the mean of all the wavefunctions to zero while keeping track of the standard deviations. In the position basis, with system, environment 1 and 2 having a standard deviations $\sigma, \eta_1 , \eta_2 $ respectively,

		\begin{align}
		\rho_s(t) &= \dfrac{1}{(2\pi)^{3/2} \sigma\eta_1\eta_2}\int dx \,d\xb\, dy_1\, dy_2\, \exp\left( -\dfrac{x^2+\xb^2}{4\sigma^2} - \dfrac{y_1^2+(y_1+(x-\xb)t)^2}{4\eta_1^2}-y\dfrac{y_2^2} {2\eta_2^2}\right)\nonumber\\&\hspace{3cm}\ket{x+y_2 t}\bra{\xb+y_2 t}\\
		&= \dfrac{1}{2\pi \sigma\eta_2}\int dx \,d\xb\, dy_2 \exp\left( -\dfrac{(x-\xb)^2 t^2}{8\eta_1^2} -\dfrac{x^2+\xb^2}{4\sigma^2} -\dfrac{y_2^2}{\eta^2} \right)\ket{x+y_2 t}\bra{\xb+y_2 t}
		\end{align}
	
		Doing a change of variables $ X=x+y_2 t, \overline{X} = \xb+y_2 t$, for which $\det(J) = 1$, and integrating out $y_2$
		\begin{align}
		\rho_s(t) &= \dfrac{1}{\sqrt{2\pi(\sigma^2+\eta_2^2t^2)}} \int dX\,d\overline{X}\,\exp\left( -\dfrac{X^2+\overline{X}^2}{4(\sigma^2+\eta^2_2 t^2)} - \dfrac{(\sigma^4 t^2 + \eta_1^2\eta_2^2 t^2 + \sigma^2 \eta_2^2 t^2)(X-\overline{X})^2}{8\sigma^2\eta_1^2(\sigma^2 +\eta_2^2 t^2)}\right)\nonumber\\&\hspace{4cm}\ket{X}\bra{\overline{X}}
		\end{align}
		This has a large $t$ behaviour of the following form
		\begin{align}
		\rho_s(t)& \simeq \dfrac{1}{\sqrt{2\pi} \eta_2 t}\int dX\,d\overline{X}\,\exp\left( -\dfrac{(X-\overline{X})^2 t^2}{8\eta_1^2} -\dfrac{(X+\overline{X})^2}{8\eta_2^2 t^2}  - \dfrac{(X-\overline{X})^2}{8\sigma^2} \right) \ket{X}\bra{\overline{X}}
		\end{align}
		The off diagonal terms die off pretty rapidly under time evolution, with the standard deviation perpendicular to the diagonal $\sim \eta_1/t$, and spreads along the diagonal linearly with time $\sim \eta_2 t$.
		Working with the same system (same wave functions) Fourier transformed in the momentum basis of the system, we get
%	\end{strip}
%\begin{strip}
%		\begin{strip}
		\begin{align}
		\hspace{-.5cm} &\rho_s(t) = \sqrt{\dfrac{2\eta_1^2 \sigma^2}{\eta_1^2 +\sigma^2 t^2}}\int dq\, d\overline{q}\, \exp\Bigg(  -\sigma^2(q^2+\overline{q}^2) \nonumber +\dfrac{2q\overline{q}t^2}{\eta_1^2 +\sigma^2 t^2} -\dfrac{(\eta_1^2\eta_2^2 -\sigma^4 +\sigma^2\eta_2^2 t^2)(q-\overline{q})^2 t^2}{2(\eta_1^2 +\sigma^2 t^2)}\Bigg)  \ket{q}\bra{\overline{q}}
		\end{align}
		which has a large time behaviour
		\begin{align}
		\rho_s(t) &\simeq \dfrac{\sqrt{2}\eta_1}{\sqrt{\pi} t } \int dq\,d\overline{q} \exp\Bigg( -\dfrac{(q-\overline{q})^2 \eta_2^2 t^2}{2}  -\dfrac{(q+\overline{q})^2\eta_1^2}{2t^2}- \dfrac{(q-\overline{q})^2\sigma^2}{2}\Bigg) \ket{q}\bra{\overline{q}}
		\end{align}
%	\end{strip}
	Here too, the off-diagonal terms vanish with standard deviation falling as $1/(\eta_2 t)$ and the diagonal spread standard deviation growing linearly as $ \sim t/\eta_1 $. It is to be noted that the parameter that controls the off-diagonal in the position space shows up as its inverse along the diagonal in the momentum space and vice-versa.

	\section{Conclusion}
	Equations (\ref{xreddensity}) and (\ref{preddensity}) implies that  $x$ and $p$ simultaneously decohere irrespective of the initial state of the system and the environment. However, the decoherence rate depends on the initial state of the environment. At $t\rightarrow \infty$, the system equilibrates, the reduced density matrix becomes time independent and  maximally mixed leading to maximization of entanglement entropy of the system. 
	Our argument crucially depends on two factors: (i) The wave functions tend to zero when $x \rightarrow \infty$ and (ii) interaction Hamiltonian involves momentum as well as position measurements. 
%	The argument (i) implies that 
The argument as it stands is not applicable for spin states as (i) is a crucial ingredient of our analysis. $p$ dependence of interaction Hamiltonian may occur in  fundamental theories or effective field theories. The form of the interaction doesn't seem unreasonable, if for instance, one takes into effect the interaction of the system with an external electromagnetic field along with any other position dependent effective interactions. Some examples are QED and light atom interaction Hamiltonians. One may also note that even though the discussion here was in terms of $x$ and $p$, one can apply the same analysis to any generalized position and conjugate momenta, fields and conjugate momenta in particular. As the fundamental reason why $\rho_s$ becomes diagonal in both $x$ and $p$ basis is the time dependent shift in the system state induced by measurement of conjugate quantities,  this feature might be preserved even in the presence of other interactions. Here, we have studied a scenario where uniform distribution results when there are no constraints on the system. It would be interesting to study the scenario where constraints are imposed dynamically on the system. Furthermore, including the effects of the system and environment which have been ignored here might be important in this context. Also, whenever one talks about equilibration, one has a length scale and time scale in mind, within which the density operator is averaged over. The dynamics at smaller scales are ignored. Such coarse graining effectively cuts off high energy processes and introduces an upper energy cutoff. These effects can be studied by introducing a regularization scheme in the form of a lattice. The behavior of reduced density matrix seems to be crucially dependent on the strength of the interaction between the system and the environment.  These issues will be studied in a future work.  
	%An open system which is described in the uniform ensemble, as we get here, maybe used as a basis to develop statistical mechanics. 
	\appendix
	\section*{Authorship contribution statement}
%	\printcredits   
	Each of the authors contributed equally to the paper.
	\section*{Acknowledgement}
	We would like to thank Sibasish Ghosh for useful discussions, suggestions on the manuscript and bringing to our notice a paper \cite{sibasish} where similar kind of effect is reported for large time behavior of a Qubit system. TPS would like to thank Anoop Varghese, Sayani Chatterjee, Y. Chaoba Devi and PNB would like to thank Parveen Kumar, Chethan Krishnan and A P Balachandran for useful discussions. PNB and TPS gratefully acknowledge support from FRG scheme (FRG/2022/PHY$\_01$
	and FRG/2022/PHY$\_02$ respectively) of National Institute of Technology Calicut.

	% Main text
	%\section{}\label{}
	
	%% Numbered list
	%% Use the style of numbering in square brackets.
	%% If nothing is used, default style will be taken.
	%%\begin{enumerate}[a)]
	%%\item 
	%%\item 
	%%\item 
	%%\end{enumerate}  
	%
	%% Unnumbered list
	%%\begin{itemize}
	%%\item 
	%%\item 
	%%\item 
	%%\end{itemize}  
	%
	%% Description list
	%%\begin{description}
	%%\item[]
	%%\item[] 
	%%\item[] 
	%%\end{description}  
	%
	%\clearpage %%Remove this from your manuscript
	%
	%% Figure
	%\begin{figure}%[]
	%  \centering
	%%    \includegraphics{}
	%    \caption{}\label{fig1}
	%\end{figure}
	%
	%
	%\begin{table}%[]
	%\caption{}\label{tbl1}
	%\begin{tabular*}{\tblwidth}{@{}LL@{}}
	%\toprule
	%  &  \\ % Table header row
	%\midrule
	% & \\
	% & \\
	% & \\
	% & \\
	%\bottomrule
	%\end{tabular*}
	%\end{table}
	
	% Uncomment and use as the case may be
	%\begin{theorem} 
	%\end{theorem}
	
	% Uncomment and use as the case may be
	%\begin{lemma} 
	%\end{lemma}

	%\section{}\label{}
	

	%\bibliographystyle{cas-model2-names}
	%
	%% Loading bibliography database
	%\bibliography{cas-refs}
	
	
	
\end{document}